\documentclass[prl,aps,amssymb,showpacs,twocolumn]{revtex4}
\usepackage{amsmath}
\usepackage{amssymb}
\usepackage{amsthm}
\usepackage{amsfonts}
\usepackage{enumerate}
\usepackage{latexsym}
\usepackage{bm}
\usepackage{graphicx}
\usepackage{subfigure}

\newcommand{\beq}{\begin{equation}}
\newcommand{\eneq}{\end{equation}}

\input{epsf}

\begin{document}

\tolerance 10000

\newcommand{\vk}{{\bf k}}


\title{Clustering Properties and Model Wavefunctions for Non-Abelian Fractional Quantum Hall Quasielectrons}

\author{B. Andrei Bernevig$^{1,2}$ and F.D.M. Haldane$^2$}

\affiliation{ } \affiliation{$^1$Princeton Center for Theoretical
Physics, Princeton, NJ 08544} \affiliation{$^2$ Department of
Physics, Princeton University, Princeton, NJ 08544}

\begin{abstract}
We present model wavefunctions for quasielectron (as opposed to
quasihole) excitations of the unitary $Z_k$ parafermion sequence
(Laughlin/Moore-Read/Read-Rezayi) of Fractional Quantum Hall states.
We uniquely define these states through two generalized clustering conditions: they
vanish when either a cluster of $k+2$ electrons is put together, or
when two clusters of $k+1$ electrons are formed at different
positions. For Abelian Fractional Quantum Hall states ($k=1$), our
construction reproduces the Jain quasielectron wavefunction, and
elucidates the difference between the Jain and Laughlin
quasielectrons. For two (or more) quasielectrons, our
states differ from those constructed using Jain's method. By adding
our quasielectrons to the Laughlin state, we obtain a hierarchy
scheme which gives rise the non-Abelian non-unitary $\nu=\frac{2}{5}$ FQH
Gaffnian state.
\end{abstract}

\date{\today}


\maketitle

The connection between conformal field theory (CFT) and
Fractional Quantum Hall (FQH) states \cite{fubini1991,moore1991} provides model wavefunctions for non-Abelian
ground states and their quasihole excitations. A central result of
the CFT-FQH connection has been the prediction that the addition of
several units of flux creates \emph{multiple} degenerate pinned quasihole states which exhibit
non-Abelian statistics. In particular, some of the Read-Rezayi (RR)\cite{read1999} series of non-Abelian states are thought to be experimentally
relevant to the $\nu=5/2$ and $\nu=12/5$ FQH
plateaus.

Despite these successes, the FQH/CFT connection has failed to produce \emph{unique}
model wavefunctions for the Laughlin
quasielectron states. This is due to the fact that, until recently, previous attempts
at introducing quasielectrons invariably necessitated using
anti-holomorphic coordinates $z^\star$ (in some form) and then
projecting to the lowest Landau level (LLL); this procedure can be done in several ways, leading to different polynomial wavefunctions. For the non-Abelian
states, quasielectron wavefunctions are not known.
Recently, several authors \cite{hansson2007} have succeeded in expressing the Jain model quasielectron
wavefunctions for the Laughlin hierarchy sequence as CFT
correlators. However, several Abelian quasielectron
models exist (due to Laughlin, Jain, Girvin, and
others), and the \emph{physical} differences between them are
not fully understood.

In this paper we provide an explicit construction of LLL quasielectron
model wavefunctions for the $Z_k$ RR sequence. The RR $Z_k$ FQH ground-states are uniquely
defined as the smallest degree polynomials that vanish when $k+1$
particles cluster together. Our purpose is to find similar \emph{physical} clustering conditions (Hamiltonians) that
uniquely define the $1$-quasielectron state. Since quasielectrons involve the removal of flux, and hence the lowering
of the total degree of the polynomial wavefunction, a
$1$-quasielectron wavefunction of the RR states can no longer vanish when $k+1$
particles come together. We find two kinds of quasielectrons:
an Abelian $1$-quasielectron wavefunction of the RR $Z_k$ sequence vanishes when $2k+1$ particles come
together and when two clusters, each of $k+1$ particles are formed
at different positions. A non-Abelian $1$-quasielectron wavefunction vanishes
when $k+2$ particles come together and when two clusters, each of
$k+1$ particles are formed at different positions. For $k=1$
(Laughlin states), the two clustering conditions are equivalent; our $1$-quasielectron states turn out to be
identical to Jain's. The clustering conditions they satisfy explain the
numerically observed energetic superiority of Jain's quasielectron over Laughlin's.
 Our many-quasielectron states differ from Jain's. A
hierarchy scheme based on the present quasielectrons gives rise to a
non-Abelian (CFT non-unitary) state for the $\nu=\frac{2}{3}$ bosonic
($\nu=\frac{2}{5}$ fermionic) Gaffnian state
\cite{simon2007} (identical to the $(k,r)=(2,3)$ Jack polynomial \cite{bernevig2007}). While Read has recently given general arguments \cite{read2007} against the idea that FQH states described by a non-unitary CFT could be gapped, we do not have a \emph{microscopic} understanding of the failure of these states, such as an identification of their conjectured \cite{read2007} gapless bulk excitation or failure to properly screen in the non-Abelian sector. Even if they do describe critical points and not phases of matter, we consider it fruitful to investigate their properties further.

\begin{figure}
\includegraphics[width=3.5in, height=1.8in]{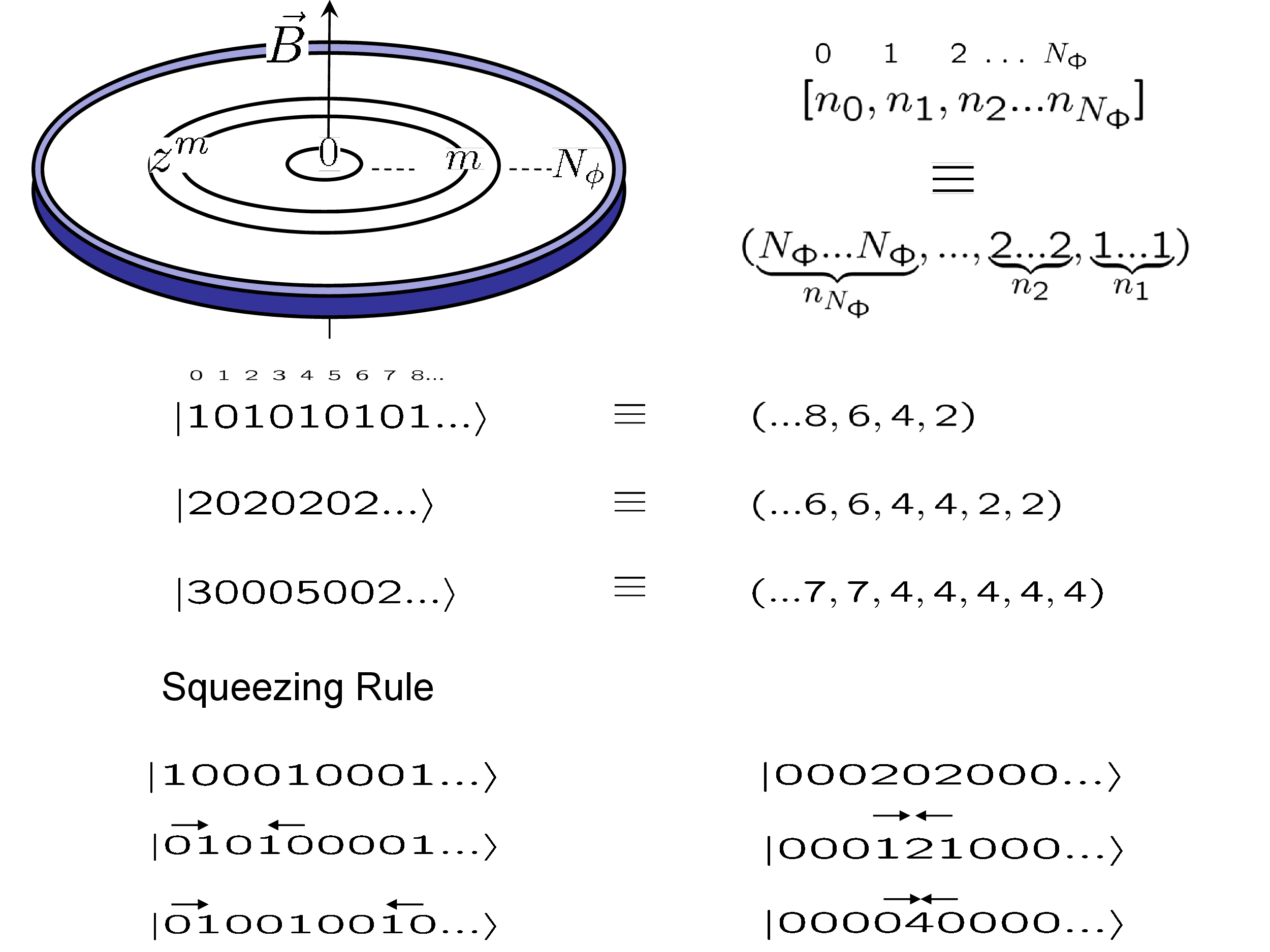}
\caption{Landau problem on a disk. Orbital occupation to monomial basis conversion and
squeezing rule examples}\label{occupation}
\end{figure}

We represent a partition $\lambda$ with length $\ell_{\lambda} \le
N$ as a (bosonic) occupation-number configuration $n(\lambda)$ =
$\{n_m(\lambda),m=0,1,2,\ldots\}$ of each of the LLL orbitals $\phi_m(z) = (2\pi m! 2^m)^{-1/2} z^m \exp(-|z|^2/4)$ with angular momentum $L_z = m \hbar$ (see
Fig[\ref{occupation}]), where, for $m > 0$, $n_m(\lambda)$ is the
multiplicity of $m$ in $\lambda$. It is useful to identify the
``dominance rule'' \cite{stanley1989} (a partial ordering of
partitions $\lambda > \mu$) with the ``squeezing
rule''\cite{sutherland1971} that connects configurations
$n(\lambda)$ $\rightarrow$ $n(\mu)$: ``squeezing'' is a two-particle
operation that moves a particle from orbital $m_1$ to $m_1'$ and
another from $m_2$ to $m_2'$, where $m_1 < m_1' \le m_2' < m_2$, and
$m_1+m_2$ = $m_1'+m_2'$; $\lambda > \mu$ if $n(\mu)$ can be derived
from $n(\lambda)$ by a sequence of ``squeezings'' (see
Fig.\ref{occupation}). An interacting LLL polynomial $P_{\lambda}$ indexed by a \emph{root partition}
$\lambda$ \cite{haldaneUCSB2006,Greiter1993} is defined as exhibiting a dominance property if it can be expanded in occupation-number non-interacting states (monomials) of orbital occupations $n(\mu)$ obtained
by squeezing on the root occupation $n(\lambda)$:
\begin{equation}
P_{\lambda} = m_\lambda + \sum_{\mu<\lambda} v_{\lambda \mu} m_\mu .
\label{dominantpolynomial}
\end{equation}
\noindent The $v_{\lambda \mu}$ are rational number
coefficients. Partitions
$\lambda$ can be classified by $\lambda_1$, their largest part. When
any $P_{\lambda}$ is expanded in monomials $m_\mu$, no orbital with $m
>\lambda_1$ is occupied. $P_\lambda$ can be interpreted as states on a sphere surrounding a monopole with
charge $N_{\Phi}=\lambda_1$\cite{haldane1983}. Uniform (ground)
states on the sphere satisfy the conditions $L^+\psi$ = 0 (highest
weight, HW) and $L^-\psi$ = 0 (lowest weight, LW) where $L^+$ =
$E_0$, and $L^-$ = $N_{\Phi}Z-E_2$, where $Z$ $\equiv$ $\sum_i z_i$,
and $E_n$ = $\sum_iz_i^n\partial/\partial z_i$. In a previous paper
\cite{bernevig2007}, we proved, by using the HW and LW conditions
that the Jack polynomials (Jacks) of root occupation
$n(\lambda^0(k,2))=[k0k0k...k0k]$ and Jack parameter
$\alpha_{k,r}=-{(k+1)}/{(r-1)}$ are the groundstate wavefunctions of the
RR $Z_k$ sequence. The RR quasihole wavefunctions are also Jacks of root occupation numbers satisfying a $(k,2)$ \cite{haldaneUCSB2006} of a more general $(k,r)$ Pauli principle
which allows no more than $k$ particles in $r$ consecutive angular momentum orbitals.  For the Jacks, the coefficients
$v_{\lambda \mu}$ are explicitly known by recursion
\cite{stanley1989}. In our
construction, we require the squeezing rule be satisfied also for the
quasi-electron states. Our root occupation number is reminiscent of the thin-torus description \cite{Bergholtz2006}; however, we generate the full interacting polynomial wavefunction and not just the non-interacting Tao-Thouless state.

Quasielectron states satisfy only the HW condition $L^+
\psi=0$ and should represent a small local perturbation of the
otherwise featureless ground state density. We now present the root occupation $n(\lambda)$ and a set of clustering
conditions which uniquely define the quasielectron wavefunctions. We start with the Abelian $1$-quasielectron added to the $\nu=\frac{k}{2}$ Jacks
$J^{-(k+1)}_{\lambda^0_{k,2}}(z_1,...z_N)$ (RR ground-states) of root occupation
$n(\lambda^0_{k,2}) = [k 0k 0k 0k...k0k]$. By analogy with the Abelian quasihole, this should be a state of total angular momentum $L=N/2$.  We add $3$ fluxes to the groundstate and obtain the occupation number: $n = [0 0 0
k0k 0k...k0k]$. The Abelian
$1-$quasielectron state is obtained by adding $2 \cdot k$ particles in the zero'th
orbital (North Pole): $n(\lambda^0_{k,1 \; qp}) = [2\cdot k 00 k
0k 0k...k0k]$. Simple counting gives us $N_\Phi=\frac{2}{k}(N-k) - 1$, the
correct flux for an $N$
particle $\nu=\frac{k}{2}$ Read-Rezayi state with an Abelian $1-$quasielectron. Away from the north
pole, the quasielectron root occupation relaxes to the bulk sequence $[k 0k...k0k]$.

The root occupation and the HW condition do not define the $1$-quasielectron polynomial wavefunction uniquely. We now search for a way to uniquely define the polynomial. In a previous paper \cite{bernevig2007A} we showed that the HW condition on
the Jacks gives an infinite set of Jacks at $\alpha=-(k+1)$ of occupation numbers
$n(\lambda^0_{k,2,s})$=$[n_0 0^{s+1}k 0k 0k0k...]$ with $n_0$ =
$(k+1)(s+1)-1$, and $s \ge 0$ a positive integer. For $s=0$ these are the RR FQH groundstates. For
$s \ge 1$, we have $n_0>k$ and hence these configurations contain an excess of charge at
the north pole, and heal in the bulk to the RR ground-state
configurations. However, as the Abelian $s$-quasielectron state in the $Z_k$ sequence should have
$N_\Phi=\frac{2}{k}(N-k) - s$,
the orbital occupation $n(\lambda^0_{k,2,s})$ contains too much charge at the north pole. To obtain the correct
$N_\Phi$, we must "subtract" $s$ particles from the zero orbital of
the occupation sequence $n(\lambda^0_{k,2,s})$=$[n_0 0^{s+1}k 0k 0k0k...]$ of the Jacks given in  \cite{bernevig2007A}, to obtain the
root occupation configuration $
n(\lambda^0_{k,s \; qp})=[k \cdot(s+1) 0^{s+1} k 0k0k...k0k]$. At the explicit, first quantized wavefunction level, this "subtraction" can be done by
symmetrization and padding of the Jack polynomial
$J^{\alpha_{k,r}}_{\lambda^0_{k,2,s}}$ \cite{bernevig2008B}, but a simpler expression will be presented shortly. Defined in this way, the $s$-quasielectron state shares
a clustering property with $J^{\alpha_{k,2}}_{\lambda^0_{k,2,s}}$ that we obtained in \cite{bernevig2007A}:
it vanishes when $s+1$ clusters of $k+1$ same-position particles are
formed. Being HW states dominated by $n(\lambda^0_{k,s \; qp})$, they also vanish when $k \cdot(s+1)+1$ particles come together at the same point as the $s+2$'s power of the difference between coordinates \cite{bernevig2007A}. The angular momentum of the Abelian $s$-quasielectron
configurations above is $l(\lambda^0_{k,s\; qp}) = L_z(\lambda^0_{k,s \; qp}) =
\frac{s}{2}N$. The above root configurations define the \emph{maximum} angular
momentum Abelian $s-$quasielectron states (bunched up at the North Pole) of the Laughlin, Read-Moore and Read-Rezayi sequence. Hence, our HW Abelian ($s=$) $1$-quasielectron state is uniquely defined as the smallest degree polynomial satisfying the clustering conditions:
\begin{eqnarray}
&P(\underbrace{z_1...z_1}_{k+1},\underbrace{z_2...z_2}_{k+1},z_{2k+3},z_{2k+4},...,z_N)
=0 \nonumber \\ &
P(\underbrace{z_1...z_1}_{2k},z_{2k+1},z_{2k+2},...,z_N) \sim
\prod_{i=2k+1}^N (z_1-z_i)^3 \label{1qpclustering}
\end{eqnarray}
\noindent For $N_\Phi=\frac{2}{k}(N-k) - 1$, the counting developed in \cite{bernevig2007A} gives exactly $N+1$
linearly independent polynomials satisfying
Eq.(\ref{1qpclustering}). They correspond to the different $l_z$'s
of the $l=\frac{N}{2}$ multiplet of states. The HW state $(l,l_z)=
(\frac{N}{2},\frac{N}{2})$ satisfies a more stringent clustering
condition than Eq.(\ref{1qpclustering}):
$P(\underbrace{z_1,...z_1}_{2k},z_{2k+1},z_{2k+2},...,z_N) =
\prod_{i=2}^N (z_1-z_i)^3
J^{-(k+1)}_{\lambda^0(k,2)}(z_{2k+1},...,z_N)$ where $n(\lambda^0(k,2))=[k0k0k...k0k]$ and $J^{-(k+1)}_{\lambda^0(k,2)}(z_{2k+1},...,z_N)$ is the RR $Z_k$ ground-state for $N-k$ particles. An alternate definition which also uniquely fixes the HW $1$-quasielectron state is requiring that it satisfies HW, dominance, and the first
clustering condition in Eq.(\ref{1qpclustering}). The second clustering condition Eq(\ref{1qpclustering}) is then automatically obeyed.

\begin{figure}
\includegraphics[width=3.7in, height=2.3in]{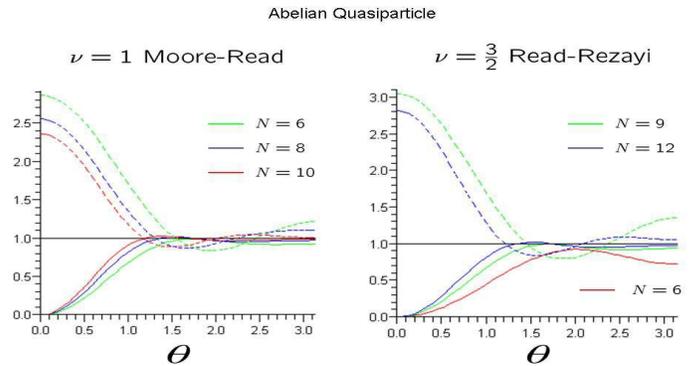}
\caption{Exact HW Abelian $1-$quasielectron (at the North Pole) (dashed) density profiles, in units of $k N/4 \pi(N-k) l^2$,
for the Moore-Read ($k=2$) and Read-Rezayi $k=3$ $N$-particle states
on the sphere. The exact LW Abelian $1-$quasihole density profiles
(solid) are also plotted for reference.}\label{densityMRRR}
\end{figure}

We now obtain explicit first quantized expressions of our states. For the Laughlin,
$(k,r)=(1,2)$, $\nu=1/2$, state we find that the $1-$quasielectron HW
wavefunction $P_{\lambda^0(1,1)}$ involves one symmetrization over a
Jack found in \cite{bernevig2007A}:
\begin{equation}
P_{\lambda^0_{1,1 \; qp}}(z_1, ...,z_N) =\text{Sym}
J^{-2}_{\lambda^0_{1,2,2}} (z_1,z_1, z_2, z_3,...z_{N})
\label{us1QP}
\end{equation}
\noindent  Model HW wavefunctions for the $s$-qp state of
\emph{maximum} angular momentum $l=s\frac{N}{2}$ are obtained by
further symmetrization over the Jacks of \cite{bernevig2007A}:
$P_{\lambda_{1,s\; qp}}(z_1,...,z_N) =\text{Sym}
J^{-2}_{\lambda^0_{1,2,s+1}} (z_1,z_1, z_2, z_2,...,z_s, z_s,
z_{s+1},z_{s+2},...,z_{N})$. For $k>1$, similar expressions can be
obtained \cite{bernevig2008B}. However, we found that our wavefunctions can be written in compact form using an operator first introduced by Jain \cite{jain1989}:
\begin{equation}
O(\partial_1,...,\partial_N, z_1,...,z_N) = Det \left(%
\begin{array}{cccc}
  \partial_1 & \partial_2 & ... & \partial_N \\
  1 & 1 & ... & 1 \\
  z_1 & z_2 & ... & z_N \\
   \vdots &\vdots &  & \vdots \\
  z_1^{N-2} & z_2^{N-2} & ... & z_N^{N-2} \\
\end{array}%
\right) \nonumber
\end{equation}
\noindent where $Det$ denotes the determinant. We find our HW Abelian $1-$quasielectron states of the RR $Z_k$ sequence, as defined by symmetrization over Jacks, are identical to:
\begin{equation}
P_{\lambda^0_{k,1 \; qp}}(z_1, ...,z_N) = \frac{1}{\Delta} O
J^{-(k+1)}_{\lambda^0_{k,2}} (z_1,...z_{N}) \label{us1QPkseq}
\end{equation}
\noindent where $\Delta =\prod_{i<j}^N (z_i-z_j)$ is the VanderMonde
determinant and $J^{-(k+1)}_{\lambda^0_{k,2}} (z_1,...z_{N})$ is the
Jack polynomial FQH ground-state of the RR $Z_k$ sequence
\cite{bernevig2007}. The right hand side of Eq.(\ref{us1QPkseq}) is
a symmetric polynomial as the determinant operator $O$
is antisymmetric in the $z_i$'s. We have checked
that $P_{\lambda^0_{k,1 \; qp}}$ in
Eq.(\ref{us1QPkseq}) exhibits a dominance property
Eq.(\ref{dominantpolynomial}) with the root occupation
$n(\lambda^0_{k,1 \; qp}) =[2\cdot k 00k0k0k0k...k0k]$, and
satisfies the clustering conditions in Eq.(\ref{1qpclustering}). The $l_z = -N/2 ...N/2$ multiplet can be obtained by successively applying the $L^-$ operator on $P_{\lambda^0_{k,1 \; qp}}$. These states also satisfy the clustering conditions in Eq.(\ref{1qpclustering}).  The density
profiles for the Read-Moore $\nu=1$ and the Read-Rezayi
$\nu=\frac{3}{2}$ quasielectron are plotted in Fig[\ref{densityMRRR}]. For $k=1$ Laughlin states, by Eq.(\ref{us1QPkseq}) our
quasielectron wavefunctions can be seen to be identical to Jain's \cite{jain1989}. Our definition of the quasielectron through the clustering conditions
Eq.(\ref{1qpclustering}) provides a physical explanation for the
numerical finding \cite{jeon2003} that Jain's quasielectron has a
lower energy than Laughlin's \cite{laughlin1983}. We found that
Laughlin's original quasielectron wavefunction \cite{laughlin1983} satisfies the second of the
clustering conditions in Eq.(\ref{1qpclustering}) but not the
first one. We have checked that Jain's quasielectron has a lower energy than Laughlin's due to the fact that it satisfies one extra clustering condition.

So far we have focused on the bosonic ($m=0$) $Z_k$ FQH
states. For integer $m\ge 1$, the Read-Rezayi sequence at filling $\nu=k/(k m+2)$
has the wavefunction $\Psi^m_{\rm RR} = \prod_{i<j} (z_i -z_j)^{m}
J^{\alpha_{k,2}}_{\lambda^0_{k,2}}$. The HW quasielectron wavefunction
is $\psi^{k,1 \; qp}_{\nu= \frac{k}{k m+2}}(z_1...z_N) =\prod_{i<j=1}^N (z_i-z_j)^{m} P_{\lambda^0_{k,1 \; qp}}(z_1, ...,z_N)
 \label{RRkmqp}$. The above construction of the quasielectron trivially generalizes to the entire $(k,r)$ Jack sequence of FQH
states introduced in \cite{bernevig2007}.

\begin{figure}
\includegraphics[width=2.7in, height=1.6in]{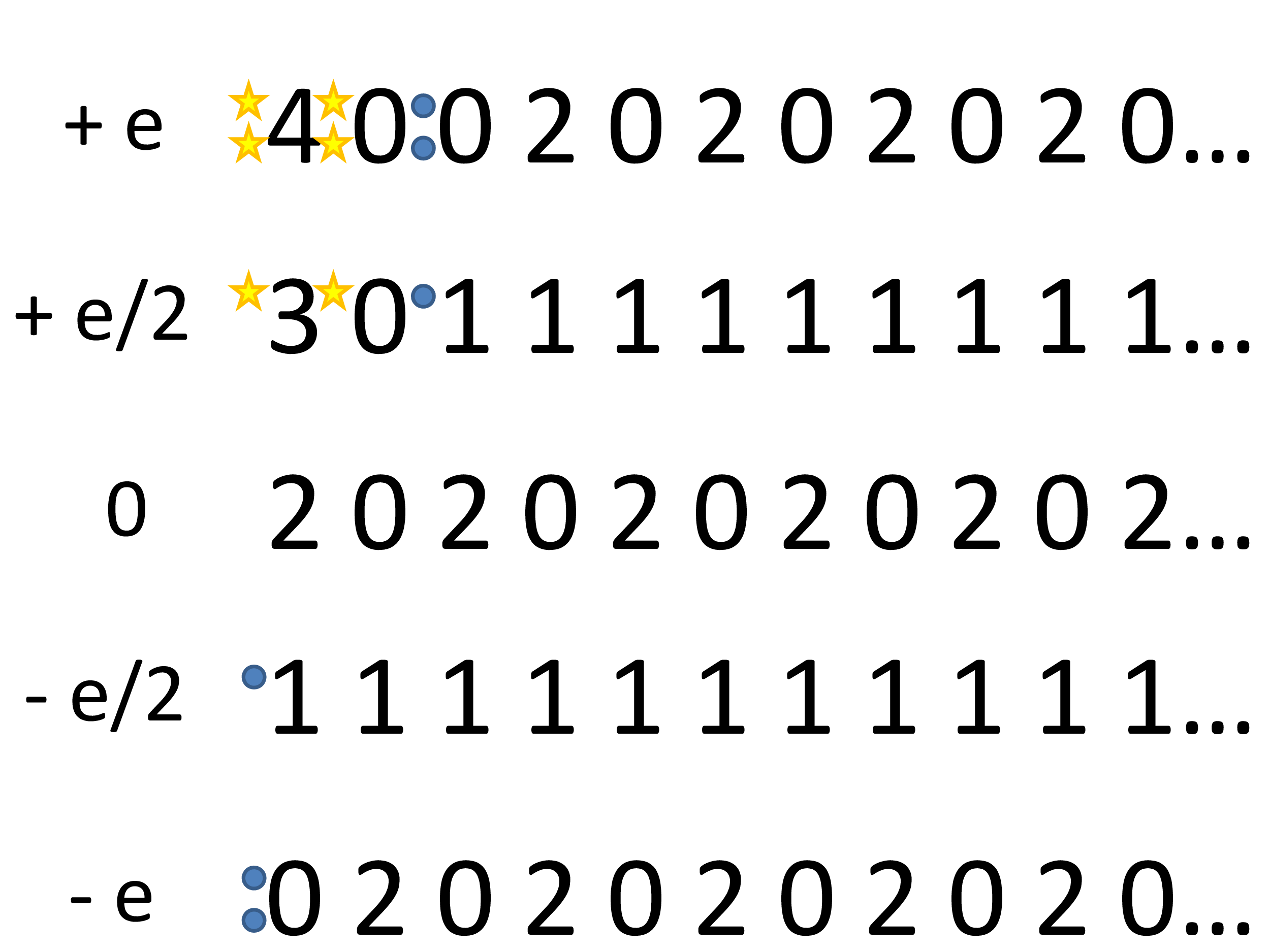}
\caption{Root occupation numbers for the Highest Weight bosonic Moore-Read ground state, Abelian (charge $-e$) quasihole, fractionalized quasihole (charge $-e/2$), Abelian (charge $e$) quasielectron, and fractionalized (charge $e/2$) quasielectron. The fractionalized quasielectron is a composite particle containing one electron, denoted here by two yellow star symbols, and a fractionalized quasihole (one blue circle)
}\label{qprootpart}
\end{figure}

We now construct the non-Abelian quasielectron
states for the RR $Z_k$ sequence. A non-Abelian fractionalized
quasielectron will always be accompanied by a non-Abelian
fractionalized quasihole, and will be composed of an electron bound to a fractionalized quasihole.  As the
Abelian quasihole has angular momentum $l=\frac{N}{2}$, each fractionalized non-Abelian quasihole (and fractionalized quasiparticle)  has $l=\frac{N}{2k}$. The basic neutral
excitation of the system is a fractionalized
$1-$quasielectron $1-$quasihole state at the same flux as the FQH
RR ground state $N_\Phi = \frac{2}{k}(N-k)$. As a fractionalized quasielectron and quasihole are distinguishable particles, angular momentum addition gives multiplets of states $l \equiv \frac{N}{2k}
\bigoplus \frac{N}{2k} = \frac{N}{k} , \frac{N}{k}-1, \frac{N}{k}-2
,...,2,1,0$ (the $l=1$ state will be missing). The HW
$l=\frac{N}{k}$ state corresponds to completely separating the
fractionalized quasielectron at the North Pole from the
fractionalized quasihole at the South Pole. It is uniquely defined
by the dominated polynomial of root occupation number $n(\lambda^0_{k,\text{1 qp  - 1 qh}}) = [k+10k-11k-11k-1...1 k-1]$, satisfying the clustering
conditions:
\begin{eqnarray}
& P(\underbrace{z_1,...,z_1}_{k+1}, \underbrace{z_2,...,z_2}_{k+1},
z_{2k+3},z_{2k+4},...,N)=0; \nonumber \\
& P(\underbrace{z_1,...,z_1}_{k+2}, z_{k+3}, z_{k+4},...,z_N)=0 \label{nonabelianclustering}
\end{eqnarray}
\noindent The HW ($l_z=l$) states of the  $l=\frac{N}{k} -1 ,...,2,0$ multiplets can be
uniquely defined by imposing HW, along with first clustering condtition in Eq.(\ref{nonabelianclustering}) on dominated polynomials with root occupation numbers:
\begin{eqnarray}
& l=\frac{N}{k};\;\;\; [k+10k-11k-11...1 k-1 1 k-1]; \nonumber \\
& l=\frac{N}{k}-1; \;\;\; [k+10k-11k-11...1 k-10k ]; \nonumber \\
& l=\frac{N}{k}-2; \;\;\; [k+10k-11k-11...1 k-1 0 k0k] ;\nonumber \\
& l=2; \;\;\; [k+10k-10k0k...k0 k0k] ;\nonumber \\
& l=0; \;\;\; [k0k0k0k...k0 k0k]  ; \label{HWnonabelian}
\end{eqnarray}
\noindent The second clustering condition in Eq.(\ref{nonabelianclustering}) is then automatically obeyed. Successive application of
the $L^-$ operator yields the $l_z = l, ...,-l$ wavefunctions, which also obey the clustering conditions in Eq.(\ref{nonabelianclustering}). We can in fact prove that our states satisfy a stronger clustering condition than in Eq.(\ref{nonabelianclustering}):
\begin{eqnarray} & P(\underbrace{z_1...z_1}_{k+1},
\underbrace{z_2...z_2}_{k}, z_{2k+2},z_{2k+3},...,z_N) \sim
(z_1-z_2)^{2 k+1} \nonumber \\ & \times \prod_{i=2k+2}^N
(z_2-z_i)^2 (z_1-z_i)^2 \label{nonabelianclustering2}
\end{eqnarray} The root occupation numbers for the Moore-Read ground-state and it's quasiparticle excitations are shown in Fig.[\ref{qprootpart}].

\begin{figure}
\includegraphics[width=3.8in, height=2.3in]{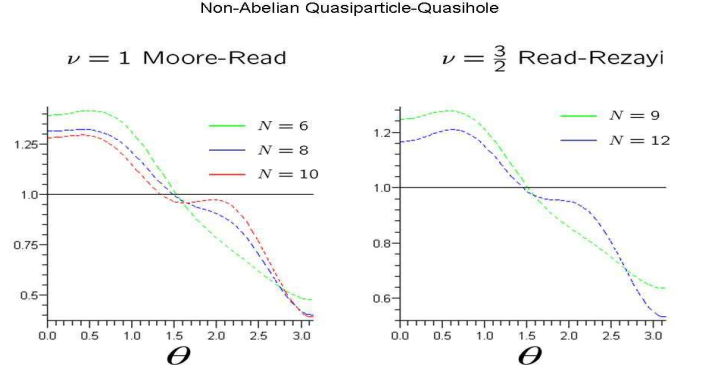}
\caption{Exact HW non-Abelian quasielectron-quasihole density
profiles,in units of $k N/4 \pi(N-k) l^2$, for the Moore-Read ($k=2$) and Read-Rezayi ($k=3$)
$N$-particle states on the sphere. The fractionalized quasielectron
is at the north pole while the fractionalized quasihole is at the
south pole. In the thermodynamic limit, the region in the middle of
the sphere at density $1$ will dominate the density function.
}\label{densityMRRRNonAbelianQPQH}
\end{figure}

Just as in the Abelian case, there are several ways to define the non-Abelian $1-$quasielectron $1-$quasihole states, which lead to the same result. Requiring HW, dominance with respect to the root occupation numbers Eq.(\ref{HWnonabelian}) and the first of the clustering conditions in Eq.(\ref{nonabelianclustering}) uniquely defines the states. The second clustering condition in Eq.(\ref{nonabelianclustering}) is then automatically satisfied. Alternatively, Eq.(\ref{nonabelianclustering2}) and the second clustering in Eq.(\ref{nonabelianclustering}) also uniquely define the Hilbert space of $1-$quasielectron $1-$quasihole states, although in this case further angular momentum projection is needed to obtain $\vec{L}$ eigenstates. For $k=1$, $z_2$ is not
different from $z_3,...,z_N$, and the non-Abelian clustering
conditions become identical to the Abelian ones (the Laughlin states support only Abelian excitations). We can ``energetically'' justify our quasielectron-quasihole wavefunctions. As they cannot
vanish when $k+1$ particles come together (this condition defines the RR $Z_k$ ground-state and pure quasiholes), the lowest ``energy'' configuration that one can
create is to require the wavefunction vanish in a $k+2$ particle cluster.

We now focus on the $s$-quasielectron states, ($s>1$) and first treat the $k=1$, $\nu=\frac{1}{2}$ Laughlin state. We have previously described the root occupation for the HW $s-$quasielectron states at
\emph{maximum} angular momentum $l=s \frac{N}{2}$, as well as their
explicit wavefunction in terms of symmetrization over a series of Jack
polynomials defined in \cite{bernevig2007A}. We now want to find the root occupation and clustering
conditions for the \emph{minimum} angular momentum $s-$quasielectron
states. Consistency arguments favor maintaining that the state vanishes when $3$
particles come together (second clustering condition for $1$-quasiparticle in Eq.(\ref{1qpclustering})). The generalization of the first clustering
condition in Eq.(\ref{1qpclustering}) is that the state vanishes when we form $s+1$ distinct clusters of $2-$particles:
\begin{eqnarray}
& P (z_1,z_1,z_2,z_2,...,z_{s+1},z_{s+1},z_{2s+3},z_{2s
+4},...,z_N)=0;
\nonumber
\\ &P (z_1,z_1,z_3,z_4,...,z_N)  \sim \prod_{i=2}^N (z_1-z_i)^3 \label{ourSqp}
\end{eqnarray}
\noindent This is the most consistent set of clustering
conditions generalizing the Eq.(\ref{1qpclustering}). For $s>1$, these states differ from Jain's. The HW state is the
minimum degree polynomial in $N$ variables satisfying Eq.(\ref{ourSqp}). We find it exhibits a
dominance property with the root configuration:
\begin{equation}
n(\lambda^0_{1,s\; qp}) =
[\underbrace{2002002...2002}_s001010101...0101]
\end{equation}
\noindent  Jain's $s-$quasielectron state is a polynomial that also
exhibits the dominance property in Eq.(\ref{dominantpolynomial}) with the
root occupation
$2010\underbrace{11011011...0110110}_{s-2}1010101...010101$, as will be shown in a future publication \cite{bernevigregnault}. It satisfies the clustering conditions \cite{bernevigregnault}:
\begin{eqnarray}
& P (z_1,z_1,z_2,z_2,...,z_{s+1},z_{s+1},z_{2s+3},z_{2s
+4},...,z_N)=0;
\nonumber
\\ & P (z_1,z_1,z_3,z_4,...,z_N)  \sim \prod_{i=2}^N (z_1-z_i)^2 \label{jainSqp}
\end{eqnarray}
\noindent but, unlike our $s-$ quasielectron state, is not uniquely defined by them. The exponent of the $s>1$ quasielectron second clustering
condition in the Jain state Eq.(\ref{jainSqp}) is different from the
exponent in the $s=1$ quasielectron Jain state Eq.(\ref{1qpclustering}). The exponent
in our clustering condition Eq.(\ref{ourSqp}) remains the same for both the $s=1$ and
$s>1$ quasielectron states. For small number of electrons with
Coulomb interaction, Jain's and our Jack quasielectron states have the same energy for $s=1$ (by virtue of being identical), and
similar energies for
$s>1$ (Monte Carlo energy for our Jack $2$ quasiparticle
bosonic state is $4.71 \pm 0.03 (\frac{e^2}{l})$ whereas Jain's
is $4.68\pm 0.01 (\frac{e^2}{l})$ for $N=6$ electrons).

A hierarchical scheme of adding our Jack quasielectrons on top of the
$\nu=1/2$ Laughlin state takes one to a \emph{non-Abelian, non-unitary} $\nu=2/3$
bosonic state. By adding $s=N/2$ Jack quasielectrons in the Laughlin
state one obtains the HW state of root occupation $[2002002...2002]$. We find this is
 the $(k,r)=(2,3)$ Jack polynomial state
\cite{bernevig2007}, initially discovered in \cite{simon2006} and called the Gaffnian. Adding  $s=N/2$ of Jain's
quasielectrons to the Laughlin state, we obtain the Abelian Jain $\nu=2/3$ state
\cite{hansson2007}. We now focus on the differences between the two
states. As the usual expression for the Jain
$\nu=2/3$ state on the plane \cite{hansson2007} does not
lead to an $\vec{L}=0$ state, we first construct it on the sphere, and then stereographically project to the
plane. We write below the root occupations of the Gaffnian and Jain's
$\nu=2/3$ states (the Jain state root configuration will be explained in a future manuscript \cite{bernevigregnault}):
\begin{eqnarray}
& Jack \; \frac{2}{3}: \;\;\; [2002002002002002...2002002002002] \\
\nonumber & Jain \; \frac{2}{3}: \;\;\;
[2010110110110110...0110110110102]
\end{eqnarray}
\noindent For a small number of particles, the Jain and the Jack
$\nu=\frac{2}{3}$ are very similar (they are identical for
$N=4$ particles). This explains their close energy and
large common overlap observed in \cite{simon2006}. From the root configuration above, we can see that Jain state is the $2-$quasielectron-$2-$
quasihole excitation of the Gaffnian Hamiltonian.

Finally, we \emph{conjecture} that the non-Abelian
$s$-quasielectron $s$-quasihole RR Hilbert space is spanned by polynomials with $N_\phi =
\frac{2}{k}(N-k)$ satisfying the clustering conditions: $1.$
$ P (\underbrace{z_1...z_1}_{k+2}, z_2, z_3,z_4...)=0$; $2.$ $P
(\underbrace{z_1...z_1}_{k+1},\underbrace{z_2...z_2}_k
z_3,z_4...)\sim \prod_{i=3}^N (z_1-z_2)^{2 k+1};$ $3.$ $P
(\underbrace{z_1...z_1}_{k+1},\underbrace{z_2...z_2}_{k+1}...\underbrace{z_{s+1}...z_{s+1}}_{k+1}
z_3,z_4...) =0$.

In this paper we have generalized the clustering conditions that
define the RR FQH ground-states and quasiholes to include the
Abelian and non-Abelian quasielectron excitations. For the Laughlin state, the Jack  $1-$quasielectron
excitations are identical to Jain's but they differ for $s>1$ quasielectrons. In particular, a hierarchy scheme based on adding Jack quasielectrons in the Laughlin state leads one to a non-Abelian non-unitary $\nu = \frac{2}{3}$ (or $\nu=\frac{2}{5}$
fermionic) state, the Gaffnian \cite{simon2006} or the Jack $(2,3)$ state \cite{bernevig2007}.

We wish to thank N. Regnault and S. Simon for numerous discussions. We also acknowledge useful discussions with T.H. Hansson, A. Karlhede,  E.J. Bergholtz, M. Hermanns, S. Viefers, E. Rezayi, E. Ardonne and P. Bonderson. This work was supported in part by the U.S. National Science
Foundation (under MRSEC Grant No. DMR-0213706 at the Princeton
Center for Complex Materials).

\emph{Note} Recently, during the editing process of this manuscript, an operator describing the non-Abelian quasiparticle state in the Moore-Read ($Z_2$) state was proposed in \cite{Hansson2008} (the full polynomial wavefunction is not given in \cite{Hansson2008} but is advertised in a longer, upcoming version of that manuscript). It would be interesting to investigate whether the two methods give identical polynomials for the Moore-Read quasiparticle state, as they do for the quasiparticle of the Laughlin state.

\end{document}